\documentstyle{article}

\begin{document}

\vspace{24pt}

\title{Theories of the Cosmological Constant}

\author{Steven Weinberg\thanks{Research
supported in part by the
Robert A. Welch
 Foundation and NSF Grants PHY 9009850 and PHY
9511632.}\\Physics Department, University of
Texas\\Austin, Texas 78712\\weinberg@physics.utexas.edu}

\date{August 29, 1996}

\maketitle

\vspace{30pt}

\noindent
{\bf Abstract} --- This is a talk given at the conference
{\it Critical Dialogues in Cosmology} at Princeton
University, June 24--27, 1996.  It gives a brief summary of
our present
theoretical understanding regarding the value of the
cosmological constant, and describes how to
calculate the probability distribution of the observed
cosmological constant in cosmological theories with a large
number of subuniverses (i. e., different expanding regions,
or different terms in the wave function of the universe) in
which this
constant takes different values.

%\begin{minipage}{4.75in}
%
%\end{minipage}
%
\vspace{100pt}

\noindent
UTTG-10-96

\vfill

\pagebreak
\setcounter{page}{1}

\section{Introduction}

The problem of the cosmological constant looks different to
astronomers and particle physicists.  Astronomers may
prefer the simplicity of a zero cosmological constant, but
they are also prepared to admit the possibility of a
cosmological constant in a range extending up to values that
would make up most of the critical density required in a
spatially flat Robertson--Walker universe.  To a particle
physicist, all the values in this observationally allowed
range seem ridiculously implausible.

   To see why, it is convenient to consider the effective
quantum field theory that takes into account only degrees of
freedom with energy below about 100 GeV, with all higher
energy radiative corrections buried in corrections to the
various parameters in the effective Lagrangian.  In this
effective field theory, the vacuum energy density that
serves as a source of the long-range gravitational field may
be written as
\begin{equation}
\rho_V=\frac{\Lambda}{8\pi G}+\frac{1}{2}\sum \hbar\omega\;,
\end{equation}
where $\Lambda$ is the cosmological constant appearing in
the Einstein field equations, and the second term symbolizes
the contribution of quantum fluctuations in the fields of
the effective field theory, cut off at particle energies
equal to 100 GeV.  Now, we know almost everything about this
effective field theory --- it is what particle physicists
call the standard model --- and we know that the quantum
fluctuations do not cancel, so that on dimensional grounds,
in units with $\hbar=c=1$, they yield
\begin{equation}
\frac{1}{2}\sum \hbar\omega\approx (100\;{\rm GeV})^4\;
\end{equation}
On the other hand, observations do not allow $\rho_V$ to be
much greater than the critical density, which in these units
is roughly $10^{-48}\;{\rm GeV}^4$.  Not to worry --- just
arrange that the Einstein term $\Lambda$ has a value for
which the two terms in Eq.~(1) cancel to fifty-six decimal
places.  This is the cosmological constant problem: to
understand this cancellation.

   Here I will consider three main directions for solving
this problem\cite{Weinberg89}:
\begin{itemize}
\item ~~~Deep Symmetries
\item ~~~Cancellation Mechanisms
\item ~~~Anthropic Constraints
\end{itemize}
By a `deep symmetry' I mean some new symmetry of an
underlying
theory, which is {\em not} an unbroken symmetry of the
effective field
theory below 100 GeV (because we know all these symmetries),
but
which nevertheless requires $\rho_V$ to vanish.  In other
contexts supersymmetry can sometimes play the role of a deep
symmetry, in the sense that some dimensionless bare
constants that are required
to vanish by supersymmetry can be shown to vanish to all
orders in perturbation theory even though supersymmetry is
spontaneously broken.  Unfortunately the vacuum density is
not a constant of this sort --- it has dimensionality
(mass)$^4$ instead of being dimensionless, and it is a
renormalized coupling rather than a bare coupling.  Recently
Witten has proposed a highly imaginative and speculative
mechanism by which
some form of supersymmetry makes $\rho_V$
vanish\cite{Witten95}.  I am grateful to the organizing
committee of this conference for giving me only 15 minutes
to talk, so that I don't have to try to explain Witten's
idea.  I turn instead to the other two approaches on my
list.

\section{Cancellation Mechanisms}

The special thing about having $\rho_V=0$ is that it makes
it possible to find spacetime-independent solutions of the
Einstein gravitational field equations.  For such solutions,
we have
\begin{equation}
\partial {\cal L}/\partial g_{\mu\nu}=0\;,
\end{equation}
where ${\cal L}$ is the Lagrangian density for constant
fields.
The problem occurs only in the trace of this equation, which
receives a contribution from $\rho_V$ which for $\rho_V\neq
0$ prevents a solution.  Many theorists have tried to get
around this difficulty by introducing a scalar field $\phi$
in such a way that the trace of $\partial {\cal L}/\partial
g_{\mu\nu}$ is proportional to $\delta {\cal L}/\delta\phi$:
\begin{equation}
g_{\mu\nu}\partial {\cal L}/\partial
g_{\mu\nu}=f(\phi)\delta {\cal L}/\delta\phi\;,
\end{equation}
with $f(\phi)$ arbitrary, except for being finite.
Where this is done, the existence of a  solution of the
field equation $\delta {\cal L}/\delta\phi=0$ for a
spacetime-independent $\phi$ implies that the trace
$g_{\mu\nu}\partial {\cal L}/\partial g_{\mu\nu}=0$ of the
Einstein field equation  for a spacetime-independent metric
is also satisfied.
The trouble is that, with these assumptions, the Lagrangian
has such a simple dependence on $\phi$ that it is not
possible to find a solution of the field equation for
$\phi$.  This is because Eq.~(4), together with the general
covariance of the action $\int d^4x\, {\cal L}$, tells us
that, when the action is stationary with respect to
variations of all other fields, it has a symmetry under the
transformations
\begin{equation}
\delta g_{\lambda\nu}=2\epsilon
g_{\lambda\nu}\;,~~~~~~\delta\phi=-\epsilon f(\phi)\;,
\end{equation}
which requires the Lagrangian density for
spacetime-independent fields $g_{\mu\nu}$ and $\phi$ to have
the form
\begin{equation}
{\cal
L}=c\;\sqrt{\det\,g}\;\exp\left(4\int^\phi\frac{d\phi'}{f(
\phi')}\right)\;,
\end{equation}
where $c$ is a constant whose value depends on the lower
limit chosen for the integral.
For $c\neq 0$, there is no solution at which this is
stationary with
respect to $\phi$.  The literature is full of proposed
solutions of the cosmological constant problem based on this
sort of spontaneous adjustment of one or more scalar fields,
but if you look at them closely, you will see  that
either they do not satisfy Eq.~(4), in which case there may
be a solution for $\phi$ but it does not imply the vanishing
of $\rho_V$, or else they do satisfy Eq.~(4), in which case
a
solution of the field equation for $\phi$ would imply a
vanishing $\rho_V$, but there is no solution of the field
equation for $\phi$.  To the best of my knowledge, no one
has found a way out of this impasse.

\section{Anthropic Considerations}

Suppose that the observed subuniverse is only one of many
subuniverses, in which $\rho_V$ takes a variety of different
values.  This is the case for instance in theories of
chaotic inflation\cite{Linde86}, in which various scalar
fields on which the vacuum energy depends take different
values in different expanding regions of space.  In a
somewhat more subtle way, this can also be the case in some
versions of quantum cosmology, where the wave function of
the universe is a superposition of terms in which $\rho_V$
takes different values, either because of the presence of
some vacuum field (like the antisymmetric tensor gauge
field  $A_{\mu\nu\lambda}$ introduced for this purpose by
Hawking\cite{Hawking84}),  or because of
wormholes, as in the work of Coleman\cite{Coleman88a}.

Some authors\cite{Hawking84}, \cite{Baum84},
\cite{Coleman88b} have argued that in quantum cosmology the
distribution of values of $\rho_V$ is very sharply peaked at
$\rho_V=0$, which would immediately solve the cosmological
constant problem.  This conclusion has been
challenged\cite{Fischler89},  and it will be assumed here
that the probability distribution of $\rho_V$ is smooth at
$\rho_V=0$, without any sharp peaks or dips.

In any theory of this general sort the measured effective
cosmological constant would be much smaller than the value
expected on dimensional grounds in elementary particle
physics, not because there is any physical principle that
makes it small in all subuniverses, but because it is only
in
the subuniverses where it is sufficiently small that there
would be anyone
to measure it.  For negative values of $\rho_V$,  this
limitation comes from the requirement that the subuniverse
must survive long enough to allow for the evolution of
life\cite{Barrow86}. For positive values of $\rho_V$ (which
are observationally more promising) the limitation comes
from the requirement that large gravitational condensations
like galaxies must be able to form before the subuniverse
begins its final exponential expansion\cite{Weinberg87}.

If you don't find this sort of anthropic explanation
palatable, consider the following fable.  You are an
astronaut, sent out to explore a randomly chosen planet
around some distant star, about which nothing is known.
Shortly before you leave you learn that because of budget
cuts, NASA has not been able to supply you with any
life-support equipment to use on the planet's surface.  You
arrive on the planet, and find to your relief that
conditions are quite tolerable --- the air is breathable,
the
temperature is about 300$^\circ$ K, and the surface gravity
is not very different from what it is on earth.  What would
you conclude about the conditions
on planets in general?  It all depends on how many
astronauts NASA has sent out.  If you are the only one then
it's reasonable to infer that tolerable conditions must be
fairly common, contrary to what planetologists would have
naturally expected.  On the other hand, if NASA has sent out
a million astronauts, then all you can conclude about the
statistics of planetary conditions is that the number of
planets with tolerable conditions is probably not much less
than one in a million --- for all you know, almost all of
the astronauts have arrived on planets that cannot support
human life.  Naturally, the only astronauts in this program
that
are in a position to think about the statistics of planetary
conditions are those like you who are lucky enough to have
landed on a planet on which they can live; the others are no
longer worrying about it.

In previous work\cite{Weinberg87} I calculated the anthropic
{\em upper bound} on the cosmological constant, which arises
from the condition that $\rho_V$ should not be so large as
to prevent the formation of  gravitational condensations on
which life could evolve.  This bound is naturally larger
than the {\em average} value
of the cosmological constant that would be measured by
typical observers, which obviously gives a better estimate
of what we
might find in our subuniverse. (Vilenkin\cite{Vilenkin95}
has advocated this point of view under the name of the
`principle of mediocrity', but did not attempt a detailed
analysis of its consequences.)  The difference is important,
because the anthropic upper bound on $\rho_V$ is
considerably larger than the largest value of $\rho_V$
allowed by observation.

I will leave the observational limits on the cosmological
constant to Dr. Fukugita's talk, but without going into
details, it seems that for a spatially flat (i.e.,
$k=0$) universe, $\rho_V$ is likely to be positive and
somewhat larger than the present mass density $\rho_0$, but
probably not larger than $3\rho_0$\cite{Ostriker96}.  On the
other hand, we know that some galaxies were
already formed at redshifts $z\approx 4$, at which time the
density of matter was larger than the present density
$\rho_0$ by a factor $(1+z)^3\approx 125$.  It therefore
seems unlikely that
a vacuum energy density much smaller than $125\rho_0$ could
have completely prevented the formation of galaxies, so the
anthropic upper bound on
$\rho_V$ cannot be much less than about $125\rho_0$, which
is much greater than the largest observationally allowed
value of $\rho_V$.

In contrast, we would expect the anthropic {\em mean} value
of
$\rho_V$ to be roughly comparable to the mass density of the
universe at the time of the greatest rate for the accretion
of matter by growing galaxies,  because it is unlikely for
$\rho_V$ to be much greater than this and there is no reason
why it should be much smaller.  (I will make this more
quantitative soon.)  Although there is evidence
that galaxy formation was well under way by a redshift
$z\approx 3$, it is quite possible that most
accretion of matter into galaxies continues to lower
redshifts, as seems to be indicated by cold dark matter
models.  In this case the anthropic mean value
$<\!\rho_V\!>$ will be considerably less than the anthropic
upper bound, and perhaps within the range allowed
observationally.

I would like to present
an illustrative example of a calculation of the whole
probability distribution of the cosmological constant that
would be measured by observers, weighted by the likelihood
that there are observers to measure it.  Instead of the very
simple model\cite{Peebles67} of galaxy formation from
spherically symmetric
pressureless fluctuations used previously\cite{Weinberg87},
here I will rely on the well-known model of Gunn and
Gott\cite{Gunn72}, which also assumes spherical symmetry and
zero pressure, but takes into account the infall of matter
from outside the initially overdense core.  This is still
far from realistic, but it will allow me to make four points
about such calculations, which should
be more generally applicable.

As shown in earlier work\cite{Weinberg87}, the condition for
a spherically symmetric fluctuation to recondense is that
\begin{equation}
\frac{500\,(\Delta\rho)^3}{729\,\rho^2}\!>\rho_V\;.
\end{equation}
where $\rho$ and $\Delta\rho$ are the average density and
the  overdensity in the fluctuation at some early initial
time, say the time of recombination.
Previously $\Delta\rho$ was assumed to be uniform within a
spherical fluctuation, but Eq.~(7) actually applies to any
sphere, with $\Delta\rho$ understood to be the spatially
averaged initial overdensity within the sphere.

Suppose
that the fluctuation at recombination consists of a finite
spherical core of volume $V$ with   positive average
overdensity
$\delta\rho$, outside of which the density takes its average
value $\rho$.  (This picture is appropriate for well
separated fluctuations.  The effects of crowding and
underdense regions will be
considered in a future paper.)  Then the average overdensity
within a larger
volume $V'$ centered on this core is $\Delta\rho=\delta \rho
V/V'$.
Assuming that Eq.~(7) is satisfied by the average
overdensity $\delta\rho$ within the core,
\begin{equation}
\left.\frac{500\,(\delta\rho)^3}{729\,\rho^2}\right|_{\rm
recomb}\!>\rho_V\;,
\end{equation}
the average overdensity $\Delta\rho$  will satisfy the
condition (7) out to a volume
$$ V_{\rm
max}=\left(\frac{500}{729\rho_V}\right)^{1/3}\rho^{-
2/3}\delta\rho\,V$$
so the total
mass that will eventually collapse is
\begin{equation}
M=\delta\rho V+\rho V_{\rm max}=V\,\delta
\rho\left[1+\left(\frac{500
\rho}{729\rho_V}\right)^{1/3}\right]\;.
\end{equation}
Once a galaxy forms, the subsequent evolution of stars and
planets and life is essentially independent of the
cosmological constant ({\em this is point 1}), so the number
of independent observers arising from a given fluctuation at
the time of recombination is proportional to the mass (9)
for those fluctuations satisfying Eq.~(8), and is otherwise
zero.  Of course, the value of the cosmological constant
might be correlated with the values of other fundamental
constants, on which the evolution of life does depend, but
the range of anthropically allowed cosmological constants is
so small compared with the natural scale (2) of densities in
elementary particle physics that within this range it is
reasonable to suppose that all other constants are fixed.
({\em This is point 2.})  The range of values of $\rho_V$
for which gravitational condensations are possible is also
so much less than the average density at the time of
recombination, that the number of fluctuations ${\cal
N}(\delta\rho,V)\,dV\,d\delta\rho$ with volume  between
$V$ and $V+dV$ and average overdensity between $\delta\rho$
and $\delta\rho+d\delta\rho$ should be nearly independent of
$\rho_V$.  ({\em This is point 3.})  If ${\cal
P}(\rho_V)\,d\rho_V$ is the {\em a priori} probability that
a random subuniverse has vacuum energy density  between
$\rho_V$ and $\rho_V+d\rho_V$, then according to the
principles of Bayesian statistics, the probability
distribution for {\em observed} values of $\rho_V$ is
\begin{eqnarray}
{\cal P}_{\rm obs}(\rho_V)&\propto& {\cal
P}(\rho_V)\int_0^\infty dV
\int_{(729\rho_V\rho^2/500)^{1/3}}^\infty d\delta\rho
\;{\cal N}(\delta\rho,V)\nonumber\\&&\times\;
V\delta \rho\left[1+\left(\frac{500
\rho}{729\rho_V}\right)^{1/3}\right]\nonumber\\
&\propto& {\cal
P}(\rho_V)\left[1+\left(\frac{500
\rho}{729\rho_V}\right)^{1/3}\right]
\int_{(729\rho_V\rho^2/500)^{1/3}}^\infty
d\delta\rho \;{\cal N}(\delta\rho)\delta\rho
\end{eqnarray}
where
\begin{equation}
{\cal N}(\delta\rho)\equiv \int_0^\infty dV\;V\, {\cal
N}(\delta\rho,V)\;.
\end{equation}
Finally, the range of values of $\rho_V$ for which
gravitational condensations are possible is so small
compared with the natural scale of densities in elementary
particle physics that within this range the {\em a priori}
probability ${\cal P}(\rho_V)$ may be taken as constant.
({\em This is point 4.})  The factor ${\cal P}(\rho_V)$ may
therefore be omitted in the probability distribution (10).
Also, all anthropically allowed values of $\rho_V$ are much
smaller than the mass density $\rho$ at recombination, so we
may neglect the $1$ in the square brackets in Eq.~(10),
which now becomes
\begin{equation}
{\cal P}_{\rm obs}(\rho_V)\propto \rho_V^{-
1/3}\int_{(729\rho_V\rho^2/500)^{1/3}}^\infty
d\delta\rho \;{\cal N}(\delta\rho)\delta\rho\;.
\end{equation}

Strictly speaking, this gives the probability distribution
only for $\rho_V>0$.  For $\rho_V<0$ and $k=0$, all mass
concentrations that are large enough to allow pressure to be
neglected will undergo gravitational collapse.  The number
of astronomers is instead limited\cite{Barrow86} for
$\rho_V<0$ by the fact that the subuniverse itself also
collapses, in a time
\begin{equation}
T(|\rho_V|)=\frac{2\pi}{3}\sqrt{\frac{3}{8\pi G
|\rho_V|}}\;.
\end{equation}
In contrast, the probability distribution for $\rho_V>0$ is
weighted by an $\rho_V$-independent factor, the average time
${\cal T}$ in which stars provide conditions favorable for
intelligent life.  The probability distribution for negative
values of $\rho_V$ is small except for values of $|\rho_V|$
that
are small enough so that $T(|\rho_V|)$ is less than or of
order ${\cal T}$.  It will be assumed here that ${\cal T}$
is very large, so that ${\cal P}_{\rm obs}(\rho_V)$ is
negligible for $\rho_V<0$ except in a small range near zero,
and may therefore be neglected in calculating the mean value
of $\rho_V$.

Using the probability distribution (12) and interchanging
the order of the integrals
over $\delta\rho$ and $\rho_V$, we easily see that the mean
value of {\em observed} values of
$\rho_V$ is
\begin{equation}
\langle\rho_V\rangle=\frac{200<\!\delta\rho^6\!>}{729<\!
\delta\rho^3\!>\rho^2}\;,
\end{equation}
with all quantities on the right-hand side evaluated at the
time of recombination, and the brackets on the right-hand
side (unlike those in
$\langle \rho_V\rangle$) indicating averages
over fluctuations:
\begin{equation}
<\!f(\delta\rho)\!>\equiv\int_0^\infty
d\delta\rho\;{\cal N}(\delta\rho)\,f(\delta\rho)\;.
\end{equation}

It remains to use astronomical observations to calculate the
fluctuation spectrum ${\cal N}(\delta\rho)$  for the density
fluctuations at recombination, which can then be used in
Eq.~(12) to calculate the probability distribution for
$\rho_V$.  Here I will just give one example of how
information about the time of formation of galaxies can put
constraints on $<\!\rho_V\!>$.  With a positive $\rho_V$,
the core of a fluctuation with average overdensity
$\delta\rho$ at recombination will collapse at a time  when
the average cosmic density $\rho_{\rm coll}$ is less than it
would be at the time of core collapse for
$\rho_V=0$:\cite{Weinberg87}
\begin{equation}
\rho_{\rm
coll}<\!\frac{500\,\delta\rho^3}{243\,\pi^2\rho^2}\;,
\end{equation}
 with $\rho$ and $\delta\rho$ on the right-hand side
evaluated at recombination.  Using this in Eq.~(14) gives a
mean vacuum density
\begin{equation}
<\!\!\rho_V\!\!>\;>\;\frac{2\pi^2\langle \rho_{\rm
coll}\rangle}{15}\;.
\end{equation}
Even if we suppose for example that core collapse occurs for
most galaxies at a redshift as low as $z\approx 1$, then
$\rho_{\rm coll}\approx 8\rho_0$, so Eq.~(19) gives
$<\!\rho_V\!>\;>\; 10\rho_0$, which exceeds current
experimantal bounds on $\rho_V$.  On the other hand, the
median value of $\rho_V$ is less than the mean value, so the
discrepancy is less than this.  Even so, it seems that most
galaxies must be formed quite late in order for the value of
$\rho_V$ in our universe to be close to the value that is
anthropically expected.

\vspace{6pt}

\centerline{*~*~*~}

At the meeting in Princeton I learned of an interesting
paper by
Efstathiou\cite{Efstathiou95}, in which he calculated the
effect of a cosmological constant on the present
number density of L$_*$ galaxies, which he took as
a measure of the distribution function    ${\cal P}_{\rm
obs}(\rho_V)$.  In this calculation he adopted a standard
cold dark matter model for matter density fluctuations, with
amplitude at long wavelengths fixed by the measured
anisotropy of the cosmic microwave background.  Efstathiou
found that for a spatially flat
universe the galaxy density falls off rapidly (say, by a
factor 10) for values of
$\rho_V$ around 7 to 9 times the present mass
density
$\rho_0$, so that $<\!\rho_V\!>/\rho_0$ should be less than
of
order 7 to 9, giving a contribution
$\Omega_0=\rho_0/(\rho_0+\rho_V)$ of matter to the total
density somewhat greater than around 0.1, which is
consistent with
lower
bounds on the present matter density.

  At first sight this seems  encouraging, but there are
a few problems with Efstathiou's calculation.    For one
thing,
as pointed out by Vilenkin\cite{Vilenkin95}, the
probability distribution of observed values of $\rho_V$ is
related to the number of galaxies (or, more accurately, the
amount of matter
in galaxies) that {\em ever} form, rather than the number
that have formed  when the age of the universe is at any
fixed value, as assumed by Efstathiou.   However, this will
not make much difference if most galaxy formation is
complete in typical subuniverses when they are as old as our
own subuniverse.  Efstathiou also encountered another
problem that is endemic to this sort of calculation.  The
cosmological parameters that can reasonably be assumed to be
uncorrelated with the cosmological constant are the
baryon--to--entropy ratio and the spectrum of density
fluctuations at
recombination, because these are presumably fixed by events
that  happened before recombination, when any anthropically
allowed cosmological constant would have been negligible.
But the only way we know about the spectrum of
density fluctuations at recombination is to use observations
of the present microwave background (or possibly the numbers
of galaxies at various redshifts), and unfortunately the
results we obtain
from this for ${\cal N}(\delta\rho)$ depend on the value of
the cosmological
constant in {\em our} subuniverse.  In calculating ${\cal
P}_{\rm obs}(\rho_V)$ one should ideally make some
assumption about the value of $\rho_V$ in  our subuniverse,
then use this value to infer a spectrum of density
fluctuations at recombination from the observed microwave
anisotropies, and then calculate the number of galaxies that
ever form as a function of $\rho_V$, with the spectrum of
density fluctuations at recombination held fixed.  Instead,
Efstathiou calculated the number of L$_*$ galaxies as a
function of $\rho_V$, with the microwave anisotropies held
fixed, which gave ${\cal P}_{\rm obs}(\rho_V)$ an
additional spurious dependence on $\rho_V$.  This problem
was known to Efstathiou, and apparently did not produce
large errors.

There is one other problem, that did have a
significant effect in Efstathiou's calculation.  He relied
on
the standard
method\cite{Press74} of calculating the evolution of density
fluctuations
using linear perturbation theory, and declaring a galaxy to
have formed when the fractional overdensity
$\Delta\rho/\rho$
reaches a value $\delta_c$, which is taken as the fractional
overdensity of the linear perturbation at a time when a
nonlinear pressureless spherically symmetric fluctuation
would recollapse to infinite density.  He took
the effective critical overdensity for spatially flat
cosmologies  as
$\delta_c=1.68/\Omega_0^{0.28}$, with $\Omega_0\equiv 1-
\rho_V/\rho_{\rm crit}$, so that $\delta_c=3.2$
for $\Omega_0=0.1$.
But numerical calculations of Martel and
Shapiro\cite{Martel96} show that for all fluctuations that
result in gravitational recollapse, $\delta_c$ is in a range
from 1.63 to 1.69.  The upper bound 1.69 is the well-known
result $\delta_c=(3/5)(3\pi/2)^{3/2}=1.6865$ for $\rho_V=0$.
The lower bound 1.63 can also be understood
analytically\cite{Weinberg87}: it is the critical
overdensity for the case where $\rho_V$ has a value that
just barely allows gravitational recollapse
\begin{equation}
(\delta_c)_{\rm
min}=\frac{2}{\sqrt{\pi}}\,\left(\frac{729}{500}\right)^{1/3
}\,
\Gamma\left(\frac{11}{6}\right)\,
\Gamma\left(\frac{2}{3}\right)=1.629\;.
\end{equation}
With $\delta_c$ always between these bounds, it is
impossible that the effective value of $\delta_c$ for any
ensemble of fluctuations could be greater than 1.69.
Overestimating $\delta_c$ biases the calculation toward late
galaxy formation, with a corresponding increased sensitivity
to relatively small values of $\rho_V$.
Efstathiou has now re-done his calculations with $\delta_c$
given the constant value 1.68, which should be a good
approximation, and, as I interpret his results,  he finds
that this change in $\delta_c$ roughly doubles the value of
$\rho_V$ at which the present density of L$_*$ galaxies
drops by a factor 10, with a corresponding reduction in the
expected value of $\Omega_0$.  It remains to be seen whether
this change in his results will lead to a conflict with
observational  bounds
on $\Omega_0$ and $\rho_V$.

At present Martel and Shapiro are carrying out a numerical
calculation of ${\cal P}_{\rm obs}$ using Eq.~(12).

I am grateful for helpful discussions with George Efstathiou
and Paul Shapiro.

\vfill\eject
\end{document}